\def\kms{\ifmmode{\hbox{km~s}^{-1}}\else{km~s$^{-1}$}\fi}
\def\la{\mathrel{\hbox{\rlap{\hbox{\lower4pt\hbox{$\sim$}}}\raise1pt\hbox{$<$}}}}
\def\ga{\mathrel{\hbox{\rlap{\hbox{\lower4pt\hbox{$\sim$}}}\raise1pt\hbox{$>$}}}}

\documentclass[12pt,b5paper]{article}

\usepackage{aaspp}
\usepackage{longtable}
\usepackage{epsfig}
\usepackage{lscape}
\setlongtables

\begin{document}

\title {\bf The Ursa Major Cluster of Galaxies. I. Cluster Definition
and Photometric Data}

\author{R. Brent Tully$^1$, Marc A.W. Verheijen$^2$, Michael J. Pierce$^3$,
Jia-Sheng Huang$^1$, and Richard J. Wainscoat$^1$}

\affil {$^1$ Institute for Astronomy, University of Hawaii, Honolulu, HI 96822}
\affil {$^2$ Kapteyn Astronomical Institute, University of Groningen} 
\affil {$^3$ Department of Astronomy, Indiana University}

\begin{abstract}
The Ursa Major Cluster has received remarkably little attention,
although it is as near as the Virgo Cluster and contains a comparable
number of HI-rich galaxies.  In this paper, criteria for group
membership are discussed and data are presented for 79 galaxies
identified with the group.  Of these, all 79 have been imaged at $B,R,I$
bands with CCDs, 70 have been imaged at $K^{\prime}$ with a HgCdTe array
detector, and 70 have been detected in the HI 21~cm line.  A complete
sample of 62 galaxies brighter than $M_B=-16.5$ is identified.  Images
and gradients in surface brightness and color are presented at a common
linear scale.  As has been seen previously, the galaxies with the
reddest global colors are reddest at the centers and get bluer at large
radii.  However, curiously, among the galaxies with the bluest global
colors there are systems with very blue cores that get redder at large
radii. 
\end{abstract}


\newpage

\section{What is the Ursa Major Cluster?}

Within a radius of 3000~\kms\ of our Galaxy there are three
moderate-sized clusters.  By happenstance, they are all at roughly the
same distance.  Two are well-known: the Virgo and Fornax clusters.
The third is remarkably poorly known: the Ursa Major Cluster.  These
three clusters are very different in their properties.  The Virgo
Cluster is the most massive, with a velocity dispersion of 715~\kms\
and a virial radius of 730~kpc (Tully 1987; see group 11-1), and contains
a mix of early-type galaxies concentrated toward a core and late-type
galaxies over a more dispersed region.  The Fornax Cluster is the most
compact, with a velocity dispersion of 434~\kms\ and a virial radius
of only 270~kpc (see group 51-1), and a majority of members are early-type
systems.  The Ursa Major Cluster is the most poorly defined, with a
velocity dispersion of only 148~\kms\ and a virial radius of 880~kpc
(see group 12-1),
and contains essentially only late-type galaxies distributed with no
particular concentration toward any center.  The total numbers of
gas-rich systems in the Ursa Major and Virgo clusters are comparable.
These group properties are summarized in Table~1.  In terms of galaxy
content, one would construct a Virgo 
Cluster with three parts Fornax and one part Ursa Major.

The Ursa Major Cluster has been difficult to define and has received
relatively little attention for two main reasons.  The first, already
stated cause for ambiguity, is the lack of concentration toward any core.
Second, the Ursa Major Cluster lies in a particularly
confusing part of the sky because it is in the plane of the Local
Supercluster at the junction of filamentary structures.  In
particular, it lies behind the long axis of the filament of galaxies
we live in, the so-called the {\it Coma-Sculptor Cloud} (Tully \& Fisher
1987; Tully 1988$a$).  Figure~1 illustrates the location of
the Ursa Major Cluster with respect to the Virgo Cluster and ourselves.

In spite of the potential for confusion,
we claim there is probably little contamination in the Ursa Major
sample that will be presented, basically because the velocity range of
the cluster is so small.
The definition of the cluster that will be
pursued here is essentially the same as in our previous work (Tully
1987; Pierce
\& Tully 1988).  However, it is radically different from the closest
correspondence in the group catalogs by Huchra \& Geller (1982; their
group No. 60) and Geller \& Huchra (1983; their group No. 94).  Their
algorithm caused them to blend much of the structure in
the filaments Tully \& Fisher called the Coma-Sculptor Cloud and
Ursa Major Cloud into one unit.  The Ursa Major Cloud contains the
Ursa Major Cluster. 
The correspondence with the catalog by de
Vaucouleurs (1975) is better but he split the cluster in two (his groups
32 and 34) and there are membership exchanges with his CVn II group
(his No. 10 = group 14-4 in Tully 1987).  Fouqu\'e et al. (1992) similarly
split the cluster in two.  Their UMa~I~S group completely overlaps
with our version of the cluster but their UMa~I~N group is drawn from
Tully's 12-1, 12-2, and 12-3 groups.  Differences will be revisited once
our definition has been described.  The more recent catalog by
Nolthenius (1993) contains the Ursa Major Cluster as group 73 and in
this case the agreement is excellent.  There is only slight exchange
with Tully's group 12-6.

Although at least Nolthenius (1993) and Tully (1987) are consistent,
the situation is still sufficiently confused that an effort at
graphical clarification is warranted.   There will be progression
through three 
figures.  Figure~2 provides an overview of the region and the problem.
The expedient of the supergalactic coordinate system is used because
in this reference frame the cluster is equatorial so rectangular
plots are almost free of distortion.  The cluster
and associated filaments are conveniently strung out along the
longitude axis in this display. 
The symbols locate all 275 galaxies in the {\it Nearby Galaxies
Catalog} (Tully 1988$a$; $NBG$ catalog) in the projected
region with $V_{\circ} < 2000$ \kms\ ($V_{\circ} = V_{helio} + 300
{\rm sin} {\ell} {\rm cos} b$).  The filled circles identify 
members of cloud 12 = Ursa Major Cloud, which contains the Ursa
Major Cluster.  The open 
squares identify members of the foreground cloud 14 = Coma-Sculptor
Cloud. 
The crosses at the edges of the figure identify the fringes of other
filamentary clouds. 

One may be discouraged by the confusion in Figure~2, particularly with
the appreciation 
that galaxies do not come with cloud labels attached.  Figure~3
provides the third dimension.  Velocities are plotted against the
supergalactic longitude
projection in this cone diagram.  The symbols make the
same associations with clouds 12 and 14.  The members of
other clouds are ignored because, as seen in Fig.~2, they are
sufficiently distinct in
supergalactic latitude.
In panel $a$, the data is
presented with no more editorial comment than implicit in the cloud
identifications.  In panel $b$, the space has been broken up into the
group identifications given by Tully (1987, 1988$a$).

Our focus is on the Ursa Major Cluster, group 12-1.  It can be seen
that the potential confusion in this case is with the entities called
12-2, 12-3, 12+6, and 14-4 (the negative group
numbers refer to units with luminosity densities above a threshold
such that they
qualify as `bound' while the positive numbers refer to
`associations' with gravitationally unbound members).  With Figure~4,
we zoom in on 
this region.  The spatial scale is increased from Fig.~2 and the
velocity window has been shaved to $400 < V_{\circ} < 1700$ \kms.

In this new figure, only the proposed members of group 12-1 (the Ursa
Major Cluster) are represented by filled circles.  The small 12-2
group is marked by boxes with inner crosses.  The
troublesome 12-3 group members are labeled with open circles and inner
crosses.  Other members
of the 12 cloud are represented by crosses.
The 14 cloud members are
still open squares; they all belong to group 14-4 in this region.

The Ursa Major Cluster (12-1) is defined in this paper by a window in
projection, 
the $7.5{\deg}$ circle
centered at $11^h 56.9^m +49^{\circ} 22^{\prime}$ ($SGL = 66.03$,
$SGB = 3.04$) superimposed on Fig.~4, and a window in velocity,
$700 < V_{helio} + 300 {\rm sin} {\ell} {\rm cos} b < 1210$ \kms.
The fussy upper velocity bound was chosen to differentiate
from the 12-3 group at a slightly higher velocity and slightly
displaced in projection.  From Fig.~4 one sees that confusion in
projection onto the $7.5{\deg}$ circle of group 12-1 = Ursa Major
Cluster comes only from groups 12-2, 12-3, and 14-4.  The first two of
these are a problem and are discussed in the next paragraph.  On the
other hand, there is a clean velocity gap below 700 km s$^{-1}$ that
separates the 14-4 group.

Why do we differ from Fouqu\'e et al. (1992)?  If the tiny 12-2 group
is ignored, then each of us agrees that the galaxies on the
supergalactic plane spanning $73.5 < SGL < 56$ and $700 < V_0 < 1700$ \kms\
lie in two groups and the difference is in the split.  There {\it
must} be at least two groups in this window because there is a strong
gradient to higher velocities in proceeding to lower supergalactic
longitudes (see Fig.~3).  We contend that the Fouqu\'e et al. split is
unsatisfactory because their UMa~I~N still contains the strong
velocity gradient with longitude and the separation from UMa~I~S is
arbitrary.  With our split between 12-1 and 12-3 groups the Ursa Major
Cluster has only
a slight velocity gradient with longitude, at a level that could
easily be real.  Obviously, though, some individual objects in the
vicinity of $SGL \sim 60$ may be misplaced between the two groups 12-1
and 12-3.
Ten galaxies within our 12-1 group window with $SGL < 61$ are most
subject to confusion.

Is it unreasonable to call this region a cluster?  The referee says
``I see no evidence for anything that even vaguely resembles a
cluster.  This region is a supercluster filament.''  Evidently, this
region is not like our common perception of a cluster.  Maybe Figure~5
helps.  Each panel illustrates an equal window of $\Delta
SGL=14{\deg}$ and $\Delta SGB=9{\deg}$, with the middle panel
centered on the cluster and the top and bottom panels shifted up and
down the supergalactic plane.  The filled histogram is associated with
the cluster.

For the moment, let us only worry about the galaxies we associate with
the putative cluster which contribute to the filled histogram.  In
this projected area similar to the dimension of the Virgo Cluster
there are a comparable number of HI-rich galaxies as in Virgo, and
there are almost enough luminous galaxies to meet the Abell richness
zero standard.  Within the full $7.5\deg$ radius window of our cluster
definition there are 30 galaxies
within $2^m$ of the third brightest galaxy in $B$ and 26 galaxies
within $2^m$ of the third brightest galaxy in $R$.  The Abell radius
of $1.5 h^{-1}$ Mpc (Abell 1958) translates to $6.5\deg$ with our
distance corresponding to $h={\rm H}_0/100=0.85$, whence only one
bright galaxy is lost to the above counts.  The Abell richness count
is thus 29 in $B$ or 25 in $R$, slightly short of the count of 30
required to qualify as richness zero.
Besides Virgo, there is no other 2~Mpc-scale region
with anything like this richness within 3000~\kms.  It can be seen
from the group parameters in Tully (1987) that $\vert$potential energy
$\vert > \vert$kinetic energy$\vert$ for this region (ie, it is bound)
if $M/L_B \ga 40 M_{\odot}/L_{\odot}$.  In this reference it is shown
that 70\% of nearby galaxies are in groups and that essentially all
groups with at least 5 members have virial $M/L_B$ values in excess of
$40 M_{\odot}/L_{\odot}$.  Hence, it is probable that the region under
discussion is bound, which is a reasonable distinction of a `cluster'
from a 'filament'.

Admittedly, the region is not cleanly distinct from a filament.  It is
perhaps arbitrary to separate the 12+6 group/association from the
cluster.  The 12-3 and 14-4 groups are almost certainly separate but
individual galaxies could easily be given erroneous group
assignments.  There is ambiguity about the stability of the cluster
because of the strong sheer caused by the proximity of the Virgo
Cluster.  In actuality, perhaps only sub-units are bound and some
parts have positive energy.

It should be clarified that the cluster definition used here serves
some needs but not others.  It gives
a list of {\it high probability} associates with the cluster.  The result is
a sample of galaxies that {\it probably share similar distances}.  The
primary original 
motivation for this work was to identify a sample that would minimize
relative distance effects in luminosity-linewidth
distance estimator relationships (Tully \& Fisher 1977).  We want to
have an unbiased sampling of a volume to a magnitude limit.
It is not a
tragedy for this purpose to loose some cluster members as long as the
process of 
elimination is random with respect to the individual galaxy
properties.  With
windows in projection and velocity,
true members that are rejected because they
lie outside the prescribed windows would not be expected to lie in any
preferred part of the luminosity-linewidth diagrams.  The only cost
of rejection is
reduced statistics.  Of course, the erroneous inclusion of
interlopers add scatter to the derived properties of the group.
Fortunately, the cluster is actually cleanly defined except at the
border with the 12-3 group.

It is to be appreciated that a window definition would be unacceptable
if our purpose is to conduct a virial analysis, for example.  The
outlying members in velocity and space contribute significantly to the
energy budget.
There are two high velocity systems (NGC~4142 and UGC~6802)
projected onto the $7.5{\deg}$ cluster that would deserve
consideration as members and the disposition of the 12-2 triplet might be
reconsidered. 

In summary, although the Ursa Major Cluster is embedded in a messy
region, the restrictive spatial and velocity window isolates a sample
of high-probability cluster members.  This procedure works in large
part because the universe is built such that there are not many
`free-floating' galaxies and the 
velocity dispersions in groups are low (Tully 1987).

\newpage

\section{Members of the Cluster}

To date, we reckognize 79 cluster members.  Their distribution on the
sky is seen in Figure~6 free of the distraction of non-cluster
projections.  All these galaxies have 
known redshifts since an appropriate velocity is a membership prerequisite
(in the absence of velocities, the potential contamination from the Local
Supercluster would 
be intolerable).  There is sample completion within limits to be
discussed but the 79 member-designates includes all known systems that
pass through the selection window.

The principal objective when the project began was to have a complete
magnitude limited sample of late-type galaxies to aid in the
calibration of luminosity-linewidth relationships.  There is
completion for galaxies of type Sab and later to a limiting magnitude
of $m_{zw} = 15.2^m$.
Among earlier types, there is completion to the CfA1 survey limit
of $m_{zw} = 14.5^m$ (Huchra et al. 1982).
Fainter galaxies than these limits have turned up in neutral hydrogen
observations of the region of the cluster.  In particular, our
collaborative study of the Ursa Major Cluster has recently involved
observations of selected fields with the Westerbork Synthesis Radio
Telescope and a few dwarf, gas-rich objects have been uncovered.

The 79 galaxies currently accepted into the cluster are identified in
Table~2.  Col. (1) provides the {\it Principal Galaxies Catalogue}
identification (de Vaucouleurs et al. 1991).  Col. (2) gives the
common name. Cols. (3-8) contains equatorial, galactic, and
supergalactic coordinates.  The equatorial coordinates have been
determined by registration of stars on the CCD images with the Space
Telescope guide star catalog and should be accurate to $1^{''}$.
Col. (9) gives a numeric morphological
type (de Vaucouleurs et al. 1991) evaluated from the CCD images
presented in this article.  Col. (10) gives the Burstein \&
Heiles (1984) galactic reddening value at $B$ band.  Cols. (11-13) give
systemic velocities (heliocentric and adjusted for motion of 
$300 {\rm sin} \ell {\rm cos} b$ \kms) and uncertainties.  Cols. (14-15) give
linewidths at 20\% of full intensities and uncertainties.  Col. (16)
gives HI integrated fluxes in units of Jy \kms.  The velocities
and HI information are accummulated from the literature.

There are flags ``$f$'' and ``$q$'' by the names of some entries in
Table~2.  There are 17 cases with ``$f$'' that are fainter than
$M_B=-16.5$, hence not part of the luminosity-limited complete
sample.  There are 10 cases with ``$q$'' at $SGL<61$ that are
questioned members because of possible confusion with groups 12-2 and 12-3.

\newpage

\section{Imaging Photometry}

Optical images have been acquired in the $B,R_{C},I_{C}$ ($C$=Cousins)
passbands for all 79 galaxies with a variety of CCD and
telescope combinations.  Early observations were with a TI~500 device
behind a focal reducer on an 0.61m telescope at Mauna Kea Observatory
in order to acquire a large field.  Later, the detector was upgraded
to a TI~800 device and, for the most part thereafter, the
focal reducer and detector were used on the University of Hawaii
2.24m telescope.  After 1992, the focal reducer was eliminated and
observations were made at the 2.24m telescope with Tektronics CCDs,
first with 
1024 pixels on a side and later with 2048 pixels on a side.  With
these various combinations, it was possible to obtain data
with a satisfactory amount of sky background in each image.

Infrared images at $K^{\prime}$ have been
obtained for 70 of these galaxies with a 256x256 HgCdTe detector using
two telescope set-ups.  The large galaxies have been observed with the
0.61m telescope which gave an $8^{\prime}$ field, while the small galaxies
have been observed with the 2.24m telescope and a $3^{\prime}$ field.
As with the optical observations, the fields are chosen large enough
to provide sky background in each image.

A log of the observational set-ups for each galaxy and all passbands
is provided in Table~3.
The observation and reduction procedures are familiar with the CCD
material and only brief comments are in order.  Standard exposures
were 3 minutes at $R,I$ and 6 minutes at $B,K^{\prime}$ on the 2.24m
telescope and 5 minutes at $R,I$ and 10 minutes at $B,K^{\prime}$ on
the 0.61m telescope.  The high surface
brightness galaxies could saturate in the central pixels with our
standard exposures designed to reach faint levels of emission and in
such cases short exposures were also taken.  The information from the
short and long exposures are combined.  


The $K^{\prime}$ observations are somewhat more complicated because of
the high and variable sky background.  The images are composites of
six dithered exposures in each case.  An exposure sequence on a target
galaxy would always be either preceded or followed by a sequence of
exposures on blank sky with identical exposures and nearly identical
inter-readout histories.  The sky stability conditions were superior
for the observing run on the 2.24m telescope.  However, the galaxies
observed with the 0.61m telescope were the high surface brightness
objects, hence were more prominent above the sky.
A side benefit of the
dithering procedure is the elimination of dead spots from bad pixels.

All the images were reduced with two separate reduction packages.  The
CCD data was initially analyzed with the GASP software described by
Pierce (1988) and the $K^{\prime}$ data was initially analyzed by
software developed by JSH and RJW.  Subsequently, the more recently
obtained raw images and all previously reduced flatfielded images were
reanalyzed within the Groningen Image Processing SYstem (GIPSY,
http://www.astro.rug.nl/$\sim$gipsy/) environment
by MV.  This reanalysis permitted all frames in all bands of each
galaxy to be treated the same.  All available frames of a particular
galaxy were aligned and image defects, foreground stars, and companion
galaxies were masked out by hand.  Cosmic ray events were removed by a
median filtering in a $3\times 3$ pixels box dragged over the entire
image.  A more detailed description of the reduction procedures will
be given by Verheijen (Ph.D. thesis in preparation).

Figure~7 (Plates~1-9) are $B$ CCD images of the 79 galaxies in the Ursa Major
Cluster observed in our program, ordered from brightest to faintest
in integrated blue light.  The first 8 plates contain the complete
sample and the fainter galaxies are shown in the ninth.  The
linear scales are the same for 
each system.  At our estimated distance for the cluster of 15.5 Mpc,
$1^{\prime\prime} = 75$ pc.

\section{Surface Brightnesses and Scale Lengths}

Our photometric analysis begins with attempts to define the
ellipticities and position angles of isophots.  The bottom panels of
Figure~8 provide plots of
axial ratio values as a function of radius from $R$-band images.  In
roughly a quarter of 
the cases the axial ratios are well constrained and hardly change with
radius, in roughly half the cases the axial ratios are adequately
constrained, and in the remaining quarter of the cases the isophotal
axial ratios are quite unstable.  In those difficult cases, the axial
ratio instabilities are attributable (with roughly equal occurances)
to (i) interactions with companions, (ii) bars, (iii) edge-on
lenticulars where disks give 
way to bulges, and (iv) intrinsic irregularities among late,
low surface brightness systems.  The horizontal lines in the panels
labeled $b/a$ of
Fig.~8 indicate the ellipticities that we ultimately associate with
the inclinations of the galaxies.
In some cases, information leading at least partially to these
inclinations is provided 
by velocity fields (NGC's 3718, 4051, 4389, UGC's 6917, 6930) or
morphological considerations (UGC's 6816, 6922, 6956).  Radially 
averaged position angles and ellipticities ($1-b/a$) are recorded in
the summary of the photometric results, Table~4.
Once position angles and ellipticities are fixed, surface brightnesses
are computed in annuli at one arcsecond intervals.
Position angles, ellipticities, and incrementation
intervals are kept the same for all four passbands.  The run of
surface brightness with radius is shown in the top panels of Figure~8
and the color differential $B-R$ is given in the middle panels.

From inspection of the luminosity profiles in Fig.~8, it is seen that
there is good agreement between the various passbands except that the
$K^{\prime}$ material is truncated $\sim 2^m$ shallower than the $B,R,I$
material.  The sky background is much worse at $K^{\prime}$.  It would
require long exposures ($\sim 10^2$ min) to reach surface brightnesses
at $K^{\prime}$ comparable to those at optical bands.  Such long
exposures are intolerable for survey programs involving hundreds or
thousands of targets. 

Inspection of the luminosity profiles also confirm that most of the
galaxies are reasonably well described by single exponential growth
curves.  This circumstance is not surprising since essentially all
the objects clearly have disk components.  Consequently, exponential
fits have been made to all the galaxies in the sample.  The fits are
characterized by two free parameters:  a central surface brightness
and a scale length.

The fits are sometimes far from perfect.  It is common for profiles to
deviate near the center.  The deviations can be in either direction
but it is more usual for there to be excess light compared with the
expectations of the exponential disk.  These situations are well known
(cf, de Vaucouleurs 1959; Kent 1985).  In cases where the profiles deviate at
small radii, the central surface brightness associated with the
exponential disk is an inward extrapolation of the fit across the main
body of the galaxy.  Growth curves can also deviate from the
exponential relation at large radii.  There can be evidence of
truncation (van der Kruit \& Searle 1981; eg, NGC~3953, UGCs 6399,
6917, 6969).  Occasionally, there is an indication of 
upward curvature as if an $r^{1/4}$ bulge component is taking over, or
perhaps the disk approximation is inappropriate (eg, NGC~4220).  Often there is
enough uncertainty in the sky subtraction that the possibility of a
deviation from an exponential fall-off at large radius is difficult to
evaluate. 

For some purposes, it is desired to have exponential disk
two-parameter fits for a complete sample but sufficient if these
characterizations are crude.  That is, even if on occasion a
description in terms of an exponential disk does not fully make sense,
the fit is a rough description that will be useful for statistical
comparisons.  It is with this motivation that the fits were made to
all galaxies of all types in the sample.

At this stage, complicated bulge-disk separations are being avoided.
Instead, we record several direct observables that are sensitive to
the relative importance of bulges.  There is computation of effective
radii containing 20\%, 
50\%, and 80\% of the total light.  A `concentration
index' can be formed out of the ratio of the 80\% radius and the 20\%
radius.  A galaxy 
with a prominent bulge contains 20\% of its light within a relatively
small radius and consequently has a large concentration index.  We
also record the {\it measured} central surface brightness (as opposed
to the extrapolated disk central surface brightness) within a radius
of $4^{''}$ of the center.  Presuming the galaxies are at the same
distance, these surface brightnesses are relative {\it metric} quantities;
ie, they are measures of light from equal volumes of space.  For the
Ursa Major sample, $4^{''}$ radius corresponds to 300 pc.

Although in this paper our intention is to present raw data and to
hold off on interpretation,
there are some curious correlations in the basic
data that deserve to be shown.  Consider Figure~9.
The left panels compare integrated colors with the disk
central surface brightness colors while the right panels compare
ratios of scale lengths with the disk central surface brightness
colors.  The top panels compare $B$ and $R$, the middle panels compare
$B$ and $I$, and the bottom panels compare $B$ and $K^{\prime}$.

From the left panels, one sees that the {\it range} of colors is
greater in the disk central surface brightnesses than in the
integrated colors.  The correlations in the data are steeper
than the $45{\deg}$ line that maps equality between the two
measures of colors.  From the right panels, one sees that, in the
progression from redder to bluer galaxies, {\it the redder scale
lengths increase in comparison with the blue scale lengths}.  It is
not surprising that scale lengths are shorter toward the red for big
galaxies as this observation is consistent with the proposition that
galaxies get bluer at larger radii (cf, de Jong 1996).  However, it
appears that 
the situation is inverted among bluer galaxies.
The bluest galaxies are particularly bluest at their centers and get
{\it redder} at larger radii.

These bluest galaxies tend to be the faintest galaxies.  There is a
rough correlation, albeit with considerable scatter, between the
tendancy to redden with radius and total magnitude.  Roughly, the
cross-over from galaxies reddening with radius (shorter exponential
scale lengths at longer wavelengths) to the inverse occurs at $M_B
\sim -17$.  Perhaps the phenomenon can be understood at the faintest
end because the centers of such galaxies are ambiguous and may tend
to be defined by the brightening caused by recent star formation.
Older stars would be more diffused.  However, it is remarkable that
the progression in properties in Fig.~9 is so tight.

\section{Isophotal vs. Total Magnitudes}

Fortunately, wide-field CCD photometry catches all but a few percent
of the total light of high surface brightness galaxies.  The common
practise is to extrapolate 
to total magnitudes with the assumption that the light at large radii
falls off in the manner of an exponential disk with central surface
brightnesses and scale lengths that can be characterized by fits to
the main body of the galaxies (Willick 1991; Courteau 1992; Mathewson,
Ford, \& Buchhorn 1992; Giovanelli et al. 1994).  Sometimes the
extrapolation is to a specific isophotal level (Schommer et al. 1993)
or sometimes both isophotal and total magnitudes are provided (Han
1992; Lu et al. 1993).

The extension of the Ursa Major sample across a wide dynamic range to
very low surface brightness galaxies provides an opportunity to study
this problem in some detail.  {\it While extrapolations to total
magnitudes are small for bright galaxies, the extrapolations become
increasingly important for fainter galaxies.}
If light profiles are approximated by
exponential decay with radius, it is theoretically anticipated and
observationally confirmed that the fraction of the light contained
within a specified isophotal level is a simple function of the disk
central surface brightness.  The total luminosity in some passband
$\lambda$ is:
\begin{equation}
L_T^{\lambda} = L_{lim}^{\lambda} + 2\pi(b/a)\Sigma_0 
\int_{x_{lim}}^{\infty} x e^{-x/h} dx
\end{equation}
where the observed luminosity within a limiting isophot is $L_{lim}$,
the axial ratio of the isophots is $b/a$, the exponential disk central
surface brightness is $\Sigma_0$ in solar units per arcsec$^2$, the
disk scale length is $h$, and the radius from the center is $x$.
Performing the integration:
\begin{equation}
L_T^{\lambda} = L_{lim}^{\lambda} - 2\pi(b/a)\Sigma_0 h^2
[(1+(x/h)) e^{-x/h}]_{x_{lim}}^{\infty}
\end{equation}
These relations can be transformed to logarithmic units, where 
$\mu_0 = -2.5 {\rm log} \Sigma_0$, so that the total magnitude is
\begin{equation}
m_T^{\lambda} = \mu_0^{\lambda} -2.5 {\rm log} 2\pi(b/a) 
-5 {\rm log} h
\end{equation}
and the magnitude within an isophot corresponding to the radius $x$ is
\begin{equation}
m_x^{\lambda} = m_T^{\lambda} - 2.5 {\rm log} [1-(1+(x/h)) e^{-x/h}]
\end{equation}
At $n$ scale lengths, $x/h=n$, the surface brightness drops by
\begin{equation}
-2.5{\rm log} e^{-n} = 1.086 n.
\end{equation}
Hence, we can specify the number of scale lengths we observe between
$\mu_0$ and $\mu_{x_{lim}}$
\begin{equation}
\Delta n = (\mu_{x_{lim}} - \mu_0)/1.086
\end{equation}
Hence, the extrapolation beyond the observed $m_x^{\lambda}$ is
\begin{equation}
\Delta m_{ext} = 2.5 {\rm log} [1-(1+\Delta n) e^{-\Delta n}].
\end{equation}
{\it The fraction of the total light above, or below, a given isophot
just depends on the number of scale lengths, $\Delta n$, down the
exponential growth curve to the specified isophot.  Hence, there is no
dependency on the scale length $h$ or the axial ratio $b/a$.}  
The formulation provided by equations (6) and (7) is a simplification
of those presented by the references in the first paragraph of this
section.  As usual, we 
require that the axial ratio is constant and the growth
curve is well described by an exponential form.  

The Ursa Major sample provides a wide enough range of central surface
brightnesses to test the extrapolation model.  While it is impossible
to know what is going on at the isophotal levels lost below the sky, it is
possible to test that the luminosity growth curves are behaving as
expected as the faintest observable levels are approached.  In
Figure~10, there 
is a comparison between observations and model with the increment to
total magnitudes over a two magnitude surface brightness interval just
above the limiting observed isophots.  The 
following isophotal levels are taken as the limits of our observations
because these levels are just slightly above the levels we can
consistently reach: $\mu_{lim}^B = 27$, $\mu_{lim}^R = 26$, 
$\mu_{lim}^I = 25.5$, and  $\mu_{lim}^{K^{\prime}} = 23.5$.  Hence,
the points plotted in Fig.~10 indicate the changes in magnitudes in the
2 mag isophotal range above these limits.  Members of the complete
sample ($M_B<-16.5$) are seen as filled circles.  The expectations of
the exponential disk growth curve model are given by the solid
curves.  The four panels give the equivalent information in the four
passbands.  There is the least scatter at $R$ because
the sky background conditions are most favorable.  The scatter is only
marginally worse at $B$ (CCD quantum efficiency is lower) and at $I$
(sky background is increased), but considerably worse at $K^{\prime}$
(sky background is much higher compared with the galaxy signal).

In fact, the model corrections are a small, but statistically
significant, amount {\it larger} than the observed magnitude
increments between 
the last two-magnitude isophotal surface brightness contours.  The
differences between passbands are not statistically significant.
Similar results are found if, say, the last one-magnitude interval is
considered instead.  Averaging over the different passbands and
different magnitude increments near but above the faint limits, it is
concluded that the model {\it overestimates} the magnitude increments
by 12\%, with $3 \sigma$ significance compared with no adjustment.
There are two plausible reasons why the exponential model might
give an overestimate: (i) there could be an additional bulge component
so additions to the disk contribute fractionally less to the total
light, and (ii) the disk may truncate at large radii (van der Kruit \&
Searle 1981).  Since these possibilities are very real, we take an
empirical approach and accept that statistically corrections should
only be 88\% of the exponential disk model corrections.  It might be
possible to do better by considering the particulars of individual
galaxies but the corrections are usually small and uncertainties
to corrections usually negligible.  There are inserts in the panels of
Fig.~10 that illustrate the adjustments that are adopted as a function
of disk central surface brightness.  Within the range of the complete
sample, adjustment to total magnitudes are $<0.12$ even at
$K^{\prime}$ and uncertainties are $<0.04$.  However, the figures warn
us that the situation rapidly becomes less favorable as one enters the
dwarf regime.  With dwarfs much of the light might lurk below the sky
cut-off, especially at $K^{\prime}$.

\newpage

\section{Tables of Photometric Results}

The directly measured optical and near-infrared photometry results are
accummulated in
Table~4.  The second column contains general information.  Row~1: a numeric
formulation of the morphological type (carried over from
Table~2); row~2: the galaxy position angle (measured east from
north); row~3 the ellipticity $\epsilon = 1-(b/a)$ where $b/a$ is the
observed 
ratio of minor axis to major axis; row~4: the radius in arcsec at the
isophotal level of 25 mag arcsec$^{-2}$ in $B$ band.

With the following columns, each row carries information for one of
the photometric bands: $B,R,I$, and $K^{\prime}$ respectively.
Col.~4: the
dates indicate when the {\it best} data was obtained for a given
object and filter.  Col.~5: isophotal magnitudes; the isophotal levels
are 27.0 in $B$, 26.0 in $R$, 25.5 in $I$, and 23.5 in $K^{\prime}$.
Col.~6: total magnitudes; the isophotal magnitudes given in the
previous column are extrapolated to infinity as described in
Section~5.  Col.~7: the mean surface brightness of the center of the
galaxy within an ellipse of axial ratio $b/a$ and major axis radius of
$4^{''}=300$ pc.  Col.~8: the extrapolated
exponential disk central surface brightness.  Col.~9: the disk
exponential scale length.  Cols. 10-12: the radii containing 20\%,
50\%, and 80\% of the light, respectively.

Parameters that can be derived from the directly measured results are
gathered for convenience into Table~5.  The magnitudes in this table
are adjusted for the effects of inclination by procedures that will
not be described in detail in this paper.  The recipes for these
corrections are based on the model described by Tully \& Fouqu\'e
(1985) except that a revised analysis has been independently carried
out in each of the passbands.  The amplitude of the correction can
be found in an individual case by comparison of the adjusted magnitude
in Table~5 with the raw total magnitude in Table~4.  For reference,
the maximum adjustment from face-on to edge-on is 0.84 mag at $B$,
0.58 mag at $R$, 0.46 mag at $I$, and 0.06 mag at $K^{\prime}$.
The absolute magnitudes are
based on a distance modulus of 30.95 (distance of 15.5 Mpc) to the
cluster.

The entries in Table~5 are the following.  Col.~1: names 
are repeated.  Col.~2: {\it top}, morphological type is repeated.;
{\it bottom}, inclinations, $i$, are calculated from ellipticities
in a standard fashion: ${\rm cos}i = \sqrt{((b/a)^2-q_0^2)/(1-q_0^2)}$
where the observed minor-to-major axial ratio is $b/a$ and the
intrinsic flattening is assumed to be $q_0=0.2$.  Cols.~3-5: {\it top},
$B,R,I,K^{\prime}$ magnitudes adjusted for internal and galactic
obscuration.; {\it bottom}, absolute magnitudes, assuming a distance
modulus of 30.95.  Cols.~7-9: {\it top}, the global colors $B-R$, $B-I$, and
$B-K^{\prime}$, respectively; {\it bottom},  
colors associated with
the exponential disk central surface brightnesses; $\mu_0^B-\mu_0^R$, 
$\mu_0^B-\mu_0^I$, and $\mu_0^B-\mu_0^{K^{\prime}}$, respectively.
Cols.~10-13: {\it top, cols. 10-12}, ratios of exponential
scale lengths between the various bands; $h_R/h_B$, $h_I/h_B$, and
$h_{K^{\prime}}/h_B$, respectively; {\it top, col.13}, a light
concentration index, $C_{82}$, formed from the ratio of the effective radii
containing 80\% and 20\% of the light.  There are only small
differences in this quantity between different filter bands.  The
values formed with 
the $K^{\prime}$ data are less stable because of higher sky noise.
For these reasons, the concentration index provided here is a straight
average of the $B,R,I$ information and does not use the $K^{\prime}$
information;
{\it bottom}, a measure of the {\it excess} light above the exponential
disk component at the center of the galaxy.  The difference is taken
between the 
exponential disk central surface brightness and the mean surface
brightness within the central $4^{''}=300$ pc radius;
$\mu_0^B-\mu_4^B$, $\mu_0^R-\mu_4^R$, $\mu_0^I-\mu_4^I$, and 
$\mu_0^{K^{\prime}}-\mu_4^{K^{\prime}}$, respectively.  
Col.~14: logarithms of the total luminosities in the
$B$ and $K^{\prime}$ bands in solar units assuming the absolute
magnitude of the sun is 5.48 at $B$ and 3.36 at $K^{\prime}$.

\section{Summary}

Although the Ursa Major Cluster lacks concentration and lies in a
confusing part of space, it is reasonably cleanly defined because it
has such a small dispersion in velocity.  We define the cluster by
windows on the plane of the sky and in velocity in order to define a
list of high-probability members.  The 79 galaxies that are accepted
lie within $7.5{\deg}$ of $\alpha = 11^h 56.9^m$, $\delta =
+49{\deg} 22^{\prime}$ and have measured velocities, $V_{helio} + 300
{\rm sin}\ell {\rm cos}b$, between 700 and 1210 \kms .  Most of the
galaxies identified with the Ursa Major Cluster are spirals with
normal gas content.  The cluster almost qualifies as an Abell richness
class 0 entity but it has not received much attention because it is so
ill-defined.  Probably the cluster is at an early evolutionary state.

We have managed to image all the cluster candidates with wide-field
CCD set-ups and almost all the objects with a HgCdTe detector.  A
complete sample of 62 galaxies brighter than $M_B = -16.5$ has been
defined.  Images are provided for all the galaxies on a common metric
scale and photometric parameters have been tabulated, including a
measure of the central light concentration,
exponential disk surface brightness zero-points and scale lengths, and
effective radii containing 20\%, 50\%, and 80\% of the light.

The data presented here has not been corrected for absorption effects
or other inclination considerations.  In parallel with these optical
observations we are involved in a program of 21 cm HI observations
with the Westerbork Synthesis Radio Telescope (Verheijen, Ph.D. thesis
in preparation).  The neutral hydrogen
observations, adjustments to the raw optical information, and analyses
of the physical properties of the sample will be addressed in
subsequent papers.  The one physical property of the sample that we
have presented at this time is the curiosity that, while redder
galaxies (typically the more luminous ones) have very red centers and
tend to get bluer at large radii, the bluer (typically smaller)
galaxies tend to do the inverse.  Galaxies with bluer centers and
which usually are globally bluer tend to get redder at large radii.

\section{Acknowledgements}

This research has been supported by NATO Collaborative Research Grant
940271 and grants from the US National Science Foundation.

\newpage

\twocolumn

\appendix

\section{Appendix: Comments on Individual Objects}

\noindent{\bf
UGC 6399:} Slightly lopsided toward the SE. 

\noindent{\bf
UGC 6446:} Patchy low surface brightness disk makes it difficult to
          determine $b/a$.  There is a bright star at the southern edge
          which somewhat contaminates the background.

\noindent{\bf
NGC 3718:} This galaxy is the largest in the cluster and has by
          far the largest disk scale length.  The galaxy is so
          peculiar that it is difficult to categorize
          morphologically. It is probably 
          strongly warped although the outer disk is kinematically well
          behaved.  An obvious dust lane crosses in front of the nucleus
          which reddens the central regions considerably.  There are
          several bright stars scattered across the disk. 

\noindent{\bf
NGC 3726:} Lopsided toward the north with a rather bright star near the
          northern edge. 

\noindent{\bf
NGC 3729:} Possibly a companion of NGC 3718 and responsible for the
          peculiar appearance of that galaxy.  It has a high surface
          brightness ring 
          surrounding a bar and a blob at the NE edge.  There is a bright
          star in front of the SW part of the disk.

\noindent{\bf
NGC 3769:} Interacting with 1135+48.  A high surface brightness ring
          surrounds a central bar.  Extended and distorted HI tidal tails
          are detected in this galaxy. 

\noindent{\bf
1135+48 :} Interacting companion of NGC 3769.  It is an irregular dwarf
          probably in the process of being tidally disrupted by NGC 3769. 
          Kinematically, it merges smoothly into the HI velocity field of
          NGC 3769. 

\noindent{\bf
NGC 3782:} A bar dominated dwarf.  The position angle and ellipticity were
          estimated from the envelope of the surrounding low surface
          brightness disk. 

\noindent{\bf
1136+46 :} HI discovered companion of NGC 3782.  Among the optically
          faintest identified cluster members. 

\noindent{\bf
1137+46 :} Same comments as for 1136+46.

\noindent{\bf
UGC 6628:} Small point-like nucleus.  Bright star in SE part.  Too little
          sky in the $K^{\prime}$-band image to allow a determination of the 
          magnitude in that band. 

\noindent{\bf
UGC 6667:} Highly flattened edge-on system suggests an intrinsic
          thickness of less than 0.2. 

\noindent{\bf
UGC 6713:} Faint.

\noindent{\bf
NGC 3870:} High surface brightness bar embedded in a featureless disk or
          envelope.  There is a small knot in the NW part of the disk. 

\noindent{\bf
NGC 3877:} Very regular spiral with a small point-like nucleus.  There is
          a very bright star just outside the field of view which
          corrupts the background somewhat.          

\noindent{\bf
UGC 6773:} Faint diffuse system without a central concentration.  There is
          a very bright star just off the frame which heavily corrupts
          the background.  As a result, the magnitudes are badly
          determined. 

\noindent{\bf
NGC 3893:} Heavily distorted outer regions due to its companion NGC 3896. 
          Two pronounced spiral arms.
          A patchy tail of debris runs off to the SE
          and reaches as far as 1 arcmin south of NGC 3896.  The adopted
          ellipticity corresponds to that of the main unperturbed disk. 
          The HI kinematics shows a strong warp. 

\noindent{\bf
NGC 3896:} Companion to NGC 3983. 

\noindent{\bf
NGC 3906:} Smooth face-on disk with a bar offset from the center.  The
          adopted ellipticity was defined by the outermost isophotes. 

\noindent{\bf
UGC 6805:} Among the smallest systems in the cluster.  It shows a double
          nucleus at a small angular separation.  High surface
          brightness with very short scale length.

\noindent{\bf
NGC 3913:} Lopsided disk with a sharp southern edge and a diffuse northern
          boundary. 

\noindent{\bf
NGC 3917:} Very regular spiral with a small nucleus.

\noindent{\bf
UGC 6816:} Patchy and irregular.

\noindent{\bf
UGC 6818:} Probably interacting with a small dwarf at its NW edge.  There
          seems to be a faint $m=1$ mode spiral arm in the western part of
          this galaxy. 

\noindent{\bf
1148+48 :} This galaxy is the smallest identified cluster member.  It
          was discovered as a Markarian galaxy (MK1460).          

\noindent{\bf
NGC 3931:} Possible elliptical galaxy.

\noindent{\bf
NGC 3928:} Faint spiral structure near nucleus.

\noindent{\bf
UGC 6840:} Bar dominated nearly face-on patchy low surface brightness
          system.  Previously classified as an edge-on when only bar
          was seen.  The kinematic
          position angle is roughly 45 degrees. 

\noindent{\bf
NGC 3924:} Faint.

\noindent{\bf
NGC 3938:} Slightly lopsided toward the north.  Small point-like nucleus. 

\noindent{\bf
NGC 3949:} Note the diffuse extended halo which surrounds this system.

\noindent{\bf
NGC 3953:} One of the largest well-formed spirals in the cluster with a
          small bar and a point-like nucleus. 

\noindent{\bf
UGC 6894:} Edge-on, late.

\noindent{\bf
NGC 3972:} Same comment as for NGC 3877 and NGC 3917.

\noindent{\bf
NGC 3982:} High surface brightness central region.  Note the filament
          along the SW edge. 

\noindent{\bf
UGC 6917:} Regular low surface brightness galaxy with a small central bar. 
          There is a bright star just west of the center. 

\noindent{\bf
NGC 3985:} There is bright compact region, offset from the center, from
          which an $m=1$ mode spiral arm emerges which can be traced over
          270 degrees. 

\noindent{\bf
UGC 6922:} There seems to be a diffuse extension toward the SW. 

\noindent{\bf
UGC 6923:} Companion to NGC 3992.  Optical warp probably
          due to the tidal influence of NGC 3992. 

\noindent{\bf
UGC 6930:} Note the small central bar and the bright star at the NW edge.

\noindent{\bf
NGC 3992:} This is the brightest and fastest rotating galaxy in the cluster. 
          It dominates three companions: UGC 6923, UGC 6940 and UGC
          6969.  It shows a 
          nice grand-design spiral structure and a prominent central
          bar. 

\noindent{\bf
NGC 3990:} Companion to N3998.  A regular S0 with a clearly visible disk
          component. 

\noindent{\bf
UGC 6940:} Companion to N3992.  Among the faintest dwarfs identified in
          the cluster. 

\noindent{\bf
NGC 3998:} Companion to N3990.  It is classified as a LINER and there is
          an HI polar ring with a position angle perpendicular to the
          optical position angle.  There is an extensive system of
          globular clusters associated with this galaxy.  Although
          classified as S0 in the RC3 and Carnegie Atlas of Galaxies, NGC
          3998 might very well be an elliptical.  

\noindent{\bf
UGC 6956:} A bar dominated dwarf with a diffuse very low surface
          brightness disk. 

\noindent{\bf
NGC 4013:} This almost perfectly edge-on system shows a prominent dust
          lane.  There is a bright foreground star close to the center. 
          There is a boxy bulge which extends far along the minor axis and
          is responsible for the outer part of the luminosity profile. 
          It is extensively studied in HI and the outer gas disk is
          strongly warped. 

\noindent{\bf
UGC 6962:} Interacting with UGC 6973.  It shows a small diffuse central bar
          and the outer regions are kinematically distorted with warp
          characteristics. 

\noindent{\bf
NGC 4010:} Note the dust patches in this edge-on system.  The NE half
          seems to be more puffed up than the SW half. 

\noindent{\bf
UGC 6969:} Dwarf companion to NGC 3992.  

\noindent{\bf
UGC 6973:} Companion of UGC 6962.  It has a very high surface brightness
          central region and an obvious dust lane in the SE part. 

\noindent{\bf
1156+46 :} This very elusive system is the faintest galaxy
          in our sample with the lowest central surface brightness.  It
          was detected at the edge of the CCD.  It seems to be resolved
          into individual stars. The gradient of surface brightness
          with radius is so shallow that the exponential scale length
          could not be defined.

\noindent{\bf
UGC 6983:} Classified as a low surface brightness galaxy.  Note the small
          but obvious central bar. 

\noindent{\bf
NGC 4026:} There is an HI filament without an optical counterpart just
          south of this edge-on S0 galaxy. 

\noindent{\bf
UGC 6992:} Faint.

\noindent{\bf
NGC 4051:} This Seyfert galaxy is clearly lopsided toward the NE.  The HI
          velocity field shows global non-circular motions. 

\noindent{\bf
NGC 4085:} Probably a companion to or in recent interaction with NGC 4088.

\noindent{\bf
NGC 4088:} A high surface brightness galaxy, among the most luminous in
          the cluster.  It shows two distorted spiral arms causing
          irregular outer isophotes. 

\noindent{\bf
UGC 7089:} Somewhat lopsided.  Probably comparable to NGC 4010 when seen
          edge-on. 

\noindent{\bf
1203+43 :} One of the faintest dwarfs in the cluster.  Probably a
          companion to UGC 7094. 

\noindent{\bf
NGC 4100:} A regular tightly wound spiral system with a small nucleus. 
          The outer region with a lower surface brightness is probably
          seen more face-on as the HI kinematics reveals a warp. 

\noindent{\bf
UGC 7094:} Dwarf.

\noindent{\bf
NGC 4102:} Classified as a LINER, this galaxy shows a bright central bar
          from which two tightly wound dusty spiral arms emerge which
          almost form a closed ring.  HI is seen in absorption against a
          bright central radio source.  Note the bright foreground star
          at the western edge. 

\noindent{\bf
NGC 4111:} An edge-on S0 with a less pronounced bulge than NGC 4026
          overall.  However the central $4^{''}$ core is the most
          luminous of any galaxy in the sample.  There
          is a very bright star just off the CCD. 

\noindent{\bf
NGC 4117:} Companion to NGC 4118.  The isophotes show some twisting probably
          due to the gravitational influence of NGC 4118. 

\noindent{\bf
NGC 4118:} Companion to NGC 4117.

\noindent{\bf
UGC 7129:} Note the small bar inside the diffuse disk.  Comparable to
          NGC 3870. 

\noindent{\bf
NGC 4138:} Note the rings of various surface brightness, apparently offset
          from the center. 

\noindent{\bf
NGC 4143:} Lenticular.

\noindent{\bf
UGC 7176:} Faint, edge-on.

\noindent{\bf
NGC 4157:} There is an obvious dust lane in the nearly edge-on spiral.

\noindent{\bf
UGC 7218:} Note the bright rim at the eastern edge of this dwarf.  The
          western side is more diffuse.  There is a foreground star at
          the western edge. 

\noindent{\bf
NGC 4183:} A nearly edge-on system with a possible warp comparable to
          NGC 4100. 

\noindent{\bf
NGC 4218:} There is a bright star just off the CCD which somewhat
          corrupts the background. 

\noindent{\bf
NGC 4217:} Note the obvious dust lane in this nearly edge-on system. 
          There are many bright stars close to this galaxy. 

\noindent{\bf
NGC 4220:} There seems to be a high surface brightness ring in this S0
          galaxy. 

\noindent{\bf
UGC 7301:} Faint, edge-on.

\noindent{\bf
UGC 7401:} There is a bright star at the eastern edge of this faint dwarf
          galaxy. 

\noindent{\bf
NGC 4346:} Lenticular.

\noindent{\bf
NGC 4389:} A bar dominated system with a diffuse extended halo.

\onecolumn


\newpage



} 
\endlandscape

\newpage

\begin{figure}
\caption{
Groups and non-group galaxies in the plane of the Local Supercluster. 
The double concentric circles are centered on the Virgo Cluster; the
inner circle defines the dimension of the cluster and the outer circle
indicates the current observed turnaround radius.  The Ursa Major
Cluster lies within the other circle near the tangent with the Virgo
infall sphere.  The Local Group lies at the origin of the plot and the
$10^{\circ}$ wedge locates the zone of obscuration where there is
incompleteness.  The three stars identify Virgo, Ursa Major, and
Virgo~W, the three clusters in the supercluster plane with ${\rm log}
L_B \ge 11.5$.  Open circles identify groups with $10.5 \le {\rm log}
L_B < 11.5$ and crosses indicate groups or individual galaxies that are
fainter. 
} 
\label{1}
\end{figure}

\begin{figure}
\caption{
Projection on the sky of all galaxies in the $NBG$ catalog with
$V_{\circ} < 2000$ \kms\ and $45{\deg} < SGL < 95{\deg}$, $-10{\deg} <
SGB < 15{\deg}$.  Filled circles: identified with Ursa Major Cloud; open
squares: identified with our local Coma-Sculptor Cloud; crosses:
identified with other structures. 
} 
\label{2}
\end{figure}

\begin{figure}
\caption{
Redshift cone diagram of the region displayed in Figure~2.  Symbols
have the same meaning.  In panel $b$ the separate groups identified in
Tully (1987) are outlined.
} 
\label{3}
\end{figure}

\begin{figure}
\caption{
Amplification of Figure~2.  Velocity limits are now $400 < V_{\circ} <
1700$ \kms.  Filled circles: group 12-1 = Ursa Major Cluster; boxes with
inner crosses: group 12-2; crosses in circles: group 12-3; crosses:
other galaxies in cloud 12; open squares: galaxies in cloud 14.  The
circle encloses the Ursa Major Cluster systems.  It has a radius of
$7.5{\deg}$ and is centered at $\alpha = 11^h 56.9^m$, $\delta =
49{\deg} 22^{\prime}$.  
} 
\label{4} 
\end{figure}

\begin{figure}
\caption{
Velocity histograms of the galaxies in three equal projected areas.  The
top panel is the histogram of 49 galaxies with $73 \leq SGL < 87$.  The
middle panel involves 100 galaxies with $59 \leq SGL < 73$.  The bottom
panel involves 28 galaxies with $45 \le SGL < 59$.  In all three cases,
the galaxies lie within the latitude limits $-2 < SGB < 7$.  Galaxies
associated with the Ursa Major Cluster contribute to the filled
histogram. 
} 
\label{5}
\end{figure}

\begin{figure}
\caption{
Projected positions of the 79 galaxies in our Ursa Major sample in
supergalactic coordinates.  The cluster is contained within a
$7.5^{\circ}$ radius.  Filled circles denote types E-S0; open circles,
types S$a$-S$cd$; crosses, types S$d$-I$m$.  Larger symbols denote
galaxies with $M_B < -19$. 
} 
\label{6}
\end{figure}

\clearpage

\begin{figure}
\caption{
$B$-band CCD images of 79 galaxies in the Ursa Major Cluster.  All
images are on a common scale: at a distance of 15.5~Mpc, 1 arcmin =
4.5~kpc.  Galaxies are ordered from brightest to faintest in integrated
blue light.  The first 8 plates contain the complete sample.  Images are
reproduced with a logarithmic grey scale. 
} 
\label{7}
\end{figure}

\begin{figure}
\caption{
Surface brightnesses, colors, and isophotal axial ratios as a function
of radius.  Surface brightness dependencies are shown in the top panels,
with data from bottom to top corresponding to $B,R,I,K^{\prime}$
respectively.  Small arrows indicate the disk central surface brightness
and exponential scale length at $B$ band.  Color dependencies at $B-R$
are shown in the central panels.  Isophotal axial ratios as a function
of radius are shown in the bottom panels.  The horizontal line indicates
the axial ratio consistent with the preferred inclination. 
} 
\label{8}
\end{figure}

\begin{figure}
\caption{
Measures of radial color change.  The left panels compare integrated
colors (horizontal) vs colors from differences between the disk
central surface brightnesses (vertical).  Colors are identical at a point
along the $45{\deg}$ line.  The right panels compare scale length
ratios with the disk central surface brightnesses.  The color
comparisons are $B-R$ on top, $B-I$ in the middle, and $B-K^{\prime}$
on the bottom.
} 
\label{9}
\end{figure}

\begin{figure}
\caption{
Magnitude increments at the radial interval of the faintest two
magnitudes of the surface brightness growth curve plotted as a
function of the disk central surface brightness.  Members of the
bright complete sample are indicated by dots and the fainter
galaxies are indicated by crosses.  The expectations of the
exponential disk growth-curve model are given by the solid curves.
In the lower part of each panel, the model extrapolations beyond the
observed limiting isophotes are shown as solid curves and the 88\%
values that are the adjustments that are actually made are shown as
dashed lines.  The four panels are for $B,R,I$ and $K^{\prime}$
respectively. 
} 
\label{10}
\end{figure}

\end{document}